\documentclass[]{aa}  

\usepackage{graphicx}
\usepackage{txfonts}
\usepackage{tablefootnote}
\usepackage[colorlinks=true,citecolor=blue]{hyperref}
\begin{document}

   \title{High-resolution, high-efficiency narrowband spectroscopy with an \textit{s-p}-phased holographic grating in double pass}


    \author{C. Farret Jentink
        \inst{1}
        \and
        F. Pepe
        \inst{1}
        \and
        C. Lovis
        \inst{1}
        \and
        C. Schwab
        \inst{2}
        \and 
        F. Wildi
        \inst{1}
        \and
        A. Clawson\inst{3}
        }

   \institute{D\'epartement d’Astronomie de l’Universit\'e de Gen\`eve, Chemin Pegasi 51b, 1290 Versoix, Switzerland,
   \and
   School of Mathematical and Physical Sciences, Macquarie University, Balaclava Road, North Ryde, NSW 2109, Australia
   \and
   Wasatch Photonics, 1305 North 1000 West, Suite 120, UT 84321 Logan, United States of America
\\
             \email{casper.farret@unige.ch}
             }

   \date{Received February 21, 2025; accepted May 15, 2025}
 
  \abstract
   {High-resolution spectroscopy (R>50,000) in astronomy is typically done with echelle-type spectrographs. The science for which these instruments have proven very effective is the detection of exoplanets through the radial velocity method, and characterizing their atmospheres. However, for atmospheric characterization, it has proven tedious to detect these signals, mostly due to sensitivity constraints. While echelle-type spectrographs provide the necessary large bandwidth for radial velocity measurements, they compromise total throughput. Additionally, the need for spectral order sorting complicates the optical design and reduces throughput further. A high spectral resolution and a limited bandpass is required to measure exoplanet atmospheric absorption from the ground. Therefore, we propose a new method to achieve very high spectral resolution with significantly higher throughput within a limited bandpass, focused on a specific spectral line or set of spectral lines of interest.}
   {We describe and test a novel method for reaching a high spectral resolution with very high unpolarized diffraction efficiency in first-order employing a tuned, high fringe-density volume phase holographic (VPH) grating in double pass. Additionally, we provide lab tests highlighting the potential of such a setup.}
   {We use a wavelength-tunable laser to measure the dispersion and diffraction efficiency of a tuned VPH grating. We compare a single-pass and double-pass setup to verify the expected results. Besides, we image the resulting spectrum to assess optical quality.}
   {We find that the VPH grating we tested can reach a diffraction-limited resolving power of >140,000 in double pass, with a peak double-pass diffraction efficiency of 79\% for unpolarized light. We tested the grating at a more modest resolution of 38,000 given sampling constraints. Based on current manufacturing abilities, we estimate double-pass diffraction efficiencies over 50\% with diffraction-limited resolving powers >200,000 should be within reach from the visible to near-infrared, where the bandwidth is limited by detector size.}
   {For specific science cases where a relatively narrow wavelength regime at (ultra-)high spectral resolution is required, a double-pass VPH setup can prove to be very efficient. As the grating operates in first order, there is no need for cross-dispersion, allowing for very high total system throughputs and overall, less complicated optics. This could bring ground-breaking science to smaller class telescopes, with relatively compact instruments, and can be of special interest for exoplanet atmospheric characterization, as these observations typically require a large amount of observing time, high signal-to-noise, and high spectral resolution.}

   \keywords{instrumentation: spectrographs, techniques: spectroscopic, planets and satellites: atmospheres, infrared: planetary systems
}
    \titlerunning{High-resolution, high-efficiency narrowband spectroscopy with an \textit{s-p}-phased holographic grating in double pass}
    \authorrunning{C. Farret Jentink et al.}
   \maketitle
%

\section{Introduction}
High-resolution spectroscopy (R>50,000) in astronomy is typically done with cross-dispersed Echelle spectrographs for the UV to near-infrared (near-IR) regime. First described by \citet{nagaoka1923}, well-known instruments are for example STIS on the \textit{Hubble Space Telescope} \citep{woodgate1998}, CRIRES \& ESPRESSO on the \textit{Very Large Telescope} \citep{kaeufl2004,pepe2010}, and HIRES on \textit{Keck} \citep{Vogt1994}. 
Echelle spectrographs combine a low groove density and high spectral order, as the maximum (diffraction-limited) spectral resolving power $R$ is given by the spectral order $M$, multiplied by the number of illuminated grating lines $N$ (Eq.~\ref{eq_res}):
\begin{equation}
    \label{eq_res}
    R = M \times N.
\end{equation}
$R$ thus depends on the size of the incident beam falling on the grating and its angle-of-incidence (AOI). It is the total optical path difference imposed by the grating that counts.

The groove density, spectral order(s), and resolving power are typically matched to satisfy the instrument requirements. Another important factor to consider here is the free spectral range (FSR). The FSR of a single order $M_i$ is defined as the spectral distance from its maximum intensity peak to the peak of the adjacent order $M_{i \pm 1}$ and is given by

\begin{equation}
    \mathrm{FSR}_i = \frac{\lambda}{M_i}.
\end{equation}

Typically, at high orders (M>20), to cover a broad wavelength range (like the full visible regime, $380 \mathrm{nm} < \lambda < 750 \mathrm{nm}$), some method is required to split these spatially overlapping orders. To do this, one typically introduces a cross-disperser -- a secondary dispersive element, like a grating or prism, with a dispersion axis perpendicular to the dispersion axis of the echelle, to separate overlapping orders \citep{nagaoka1923}. As an echelle setup relies on additional optics to separate overlapping spectral orders, the total throughput of the system suffers and overall complexity increases. Besides, given echelle gratings are mechanically ruled, they are prone to suffering from manufacturing errors. For example, they can produce significant amounts of stray light because of random errors on the grating surface and can suffer from optical ghosts as a result of periodic ruling errors \citep{gao2021}. Large echelle gratings are also challenging to manufacture, and hence commercial options are limited.

A different type of grating that has gained popularity in the world of astronomical instrumentation is the volume phase holographic (VPH) grating \citep{barden1998, arns1999}. These immersed gratings have become popular because they:
\begin{itemize}
    \item Are easily tunable in groove density, size, and glass type for specific applications.
    \item Provide very high first-order diffraction efficiencies.
    \item Barely suffer from periodic or random errors, eliminating ghosts and greatly reducing scattered light.
    \item Are well protected due to their immersion in between two glass plates, allowing for easy handling and cleaning.
\end{itemize}
In a direct comparison, VPH gratings typically surpass echelle gratings in terms of efficiency, scattered light performance, and spectral resolution \citep{kielkopf1981, farretjentink2023}. VPH gratings generally operate in first order. According to Eq.~\ref{eq_res}, achieving high spectral resolution requires a greater number of illuminated lines on the grating to match the resolutions provided by echelle gratings. Thanks to their manufacturing process, VPH gratings can accommodate this need, as they can be produced with up to >6000 lines/mm \citep{barden1998}, significantly exceeding the line densities of typical echelle gratings.

However, as the groove density increases, the angle of incidence (AOI) required to achieve maximum efficiency in the first spectral order also changes. At very high groove densities, where the AOI exceeds $36^\circ$, VPH gratings become significantly less efficient for light in the \textit{s}-polarized state \citep{Baldry2004}. Note that for the remainder of this paper we adopt the definition in which the \textit{s}-polarization has an electric vector that is perpendicular to the grating lines, similar to \citet{Baldry2004}.

For measurements of unpolarized light, which are common in astronomy, the efficiency advantages of VPH gratings can be less apparent when compared to echelle gratings in high spectral resolution configurations. However, to achieve high first-order diffraction efficiency for unpolarized light at high groove densities, one can optimize the grating thickness and refractive index modulation to align the efficiency curves of \textit{s}- and \textit{p}-polarized light within a narrow wavelength range. This technique was initially described by \citet{Dickson1994}, and such specially optimized VPH gratings are often referred to as \textit{Dickson} gratings.

To achieve very high spectral resolution while maintaining high throughput, a new instrument named NIGHT \citep{farretjentink2023} will utilize this type of grating in a double-pass configuration. As part of the development program for this instrument, we present, to our knowledge for the first time, a high-efficiency, diffraction-limited experimental setup employing a single Dickson grating in a double-pass arrangement. We also compare our experimental results with theoretical predictions and briefly discuss the potential applications of this technology in astronomical instrumentation.

\section{Theory}
\label{sect:theory}

   \begin{table*}[!h]
      \caption[]{Examples of astronomical spectrographs that use a VPH grating as the main disperser. The spectrographs are sorted according to resolving power.}
         \label{tab:spectrographs}
         \begin{tabular}{llllll}
            \hline
            \noalign{\smallskip}
            Instrument & max $R$ & Density [$\ell$/mm] & AOI [$^\circ$]$^a$ & Aperture$^b$ [mm $\times$ mm] & Reference\\
            \noalign{\smallskip}
            \hline
            \noalign{\smallskip}
            NIGHT$^c$ & 75,000 & 1407 & 49.6 & 190 $\times$ 130 & \citet{farretjentink2023}\\
            HERMES & 50,000 & 3827 & 67 & 550 $\times$ 220 & \citet{sheinis2014}\\
            WIYN & 25,000 & 3300 & 65 & 480 $\times$ 210 & \citet{Bershady2008}\\
            APOGEE & 22,500 & 1009.3 & 54 & 475 $\times$ 290 & \citet{arns2010}\\
            WEAVE & 25,000 & 2500--3500 & 54 & 373 $\times$ 227 & \citet{bianco2018}\\
            MOONS & 18,300 & 1056 & 53 & 2 (mosaic) $\times$ 280 $\times$ 290 & Ernesto Oliva (co-PI MOONS)\\
            AAOmega & 10,000 & 1700 & 47 & > 260 $\times$ 190 & \citet{smith2004, sharp2006}\\
            6dF/RAVE & 8,000 & 1700 & 47 & > 260 $\times$ 190 & \citet{Saunders2001,Baldry2004}\\
            DESI & 5,500 & 1157.4 & 10-20 & 180 $\times$ 160 & \citet{edelstein2018,ishikawa2018}\\
            \noalign{\smallskip}
            \hline
         \end{tabular}
         \footnotesize{$^a$ The AOI does not specifically refer to the incidence angle on the dichromated material. The values in the table were taken from the literature and typically refer to the incidence angle on the glass substrate. However, the incidence angle on the gelatin will depend on the refractive index of the glass substrate. This information is typically not available in the literature. 
         $^b$ Here we only define the slightly ambiguous term aperture. For most instruments, the literature does not make a clear distinction between the substrate size and clear aperture.
         $^c$ For NIGHT, a VPH in double pass as described in this paper is proposed. The line density, AOI, and aperture are thus more modest.}\\
   \end{table*}

\subsection{VPH gratings}
\begin{figure}
   \centering
   \includegraphics[width=\columnwidth]{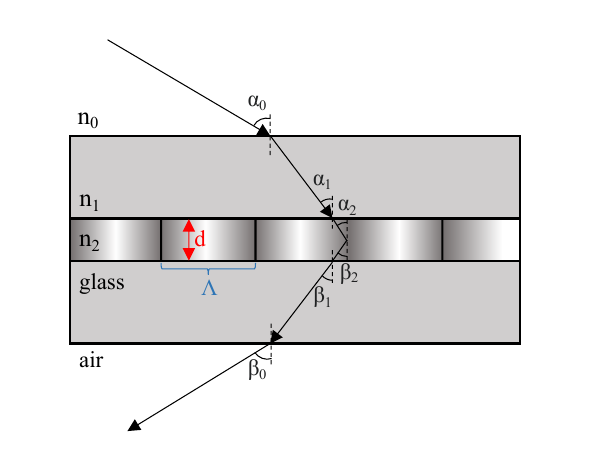}
      \caption{A diagram of a VPH grating. The dichromated gelatin layer of refractive index $\mathrm{n_2}$ and index modulation $\Delta n_g$ is immersed in two glass plates of refractive index $\mathrm{n_1}$. $d$ is the thickness of the DCG layer and $\Lambda$ the seperation between fringes. For a VPH grating in Littrow condition $\alpha_i = \beta_i$.
              }
         \label{fig:vph_diag}
   \end{figure}

All diffraction gratings operate in notably similar manners. When light strikes a grating at a specific angle of incidence $\alpha$, it creates a phase delay as it reflects off or passes through various grooves, fringes, or lines comprising the grating at different locations. Consequently, the resulting wavefronts positively interfere at distinct wavelengths for various angular deviations in the outgoing beam. This leads to the reflection or transmission of different wavelengths at different angles. The mathematical description of the interference of wavefronts leads to the Grating Equation \citep[e.g.,][]{Gover2005}:
\begin{equation}
    \label{eq:grating}
    m\rho\lambda = n_1(\sin\alpha+\sin\beta),
\end{equation}
for spectral order $m$, line-density $\rho$, wavelength $\lambda$, incidence angle $\alpha$ and diffraction angle $\beta$. \citet{Gover2005} also show that the derivative of this equation leads to the well-known equation of angular dispersion:
\begin{equation}
    \label{eq:dispersion}
    \frac{\mathrm{d}\beta}{\mathrm{d}\lambda}=\frac{m \cdot \rho}{\cos\beta},
\end{equation}
assuming the grating resides in vacuum and $n_1=1$. Now that we understand the basics of grating theory, let us take a look at VPH gratings in particular. \citet{Arns1995} \& \citet{barden1998} described the basic principles of VPH gratings and showed their potential for usage in astronomical spectrographs. A basic diagram of a VPH grating can be found in Fig.~\ref{fig:vph_diag}. A VPH is in essence built up of three layers. A central layer of dichromated material, typically dichromated gelatin (DCG), is merged between two plane glass substrates \citep{shankoff1968,barden2000, Baldry2004}. The DCG layer has a varying refractive index, acting as fringes on which incoming light can reflect. DCG has a refractive index of $\sim$1.5 and the index modulation can be tuned to values between 0.02 and 0.10 \citep{barden2000}. This modulation can typically be described by a sinusoidal variation and is applied to the DCG by holographic illumination with a wave pattern produced by fringes from a laser source interference pattern \citep{barden2000}. This way of production allows for great flexibility in the manufacturing process. We list some examples of VPH gratings in high-resolution astronomical spectrographs with their respective properties and resolving powers in Table~\ref{tab:spectrographs}. Note that some existing spectrographs achieve a very high resolving power, for example, the HERMES instrument \citep{sheinis2014}. However, it is important to note that HERMES is only sensitive in \textit{p}-polarized light \citep{heijmans2011,sheinis2014}. For HERMES, a double-pass VPH with prisms acting as reflectors was considered but ultimately discarded due to tight tolerances and limited bandwidth \citep{barden2008}, also see Sect.~\ref{sect:tolerances}.

\subsection{Diffraction efficiencies of VPH gratings}
Much literature has been written about VPH grating theory. One of the most cited examples is \citet{Kogelnik1969}, who provides an approximation for the first-order diffraction efficiencies in both polarizations. This equation is given by
\begin{equation}
    \label{eq:kogelnik}
    \eta = \frac{1}{2}\sin^2 \left( \frac{\pi \Delta n_g d}{\lambda \cos\alpha_{2}} \right) +  \frac{1}{2}\sin^2 \left( \frac{\pi \Delta n_g d}{\lambda \cos\alpha_{2}} \cos(2\alpha_{2}) \right),
\end{equation}

where the first term is for \textit{s}-polarized light, and the second term is for \textit{p}-polarized light. In Eq.~(\ref{eq:kogelnik}), $\Delta n_g$ refers to the index modulation of the grating, $d$ the thickness of the grating (excluding substrate), $\alpha_{i}$ the angle of incidence and $\lambda$ the wavelength. \citet{Baldry2004} also describe in detail how a VPH grating can be optimized to become a Dickson grating. In summary, one tunes the thickness and index modulation to maximize efficiency while forcing equality of both terms in Eq.~(\ref{eq:kogelnik}). A requirement is that this happens at or very near the Bragg angle, allowing for reflection by the grating fringes at very high diffraction efficiency (see Fig.~\ref{fig:vph_diag}). The Bragg condition is given by
\begin{equation}
    \label{eq:Bragg}
    \frac{m\lambda}{n_2} = 2 \Lambda \mathrm{sin} \alpha_2.
\end{equation}

\begin{figure}
   \centering
   \includegraphics[width=\columnwidth]{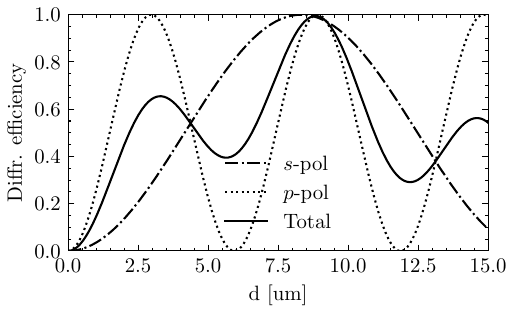}
      \caption{An example of the Kogelnik efficiency curves for \textit{s} and \textit{p}-polarization states. $\Delta n_g$ is 0.145 and $\mathrm{n_2}$ is 1.3. This grating is tuned at a DCG layer thickness of $9 \mathrm{\mu m}$ for a wavelength of 1040nm with 1400 fringes/mm and a (close to Bragg) incidence angle of $\alpha_2 = 34.55^{\circ}$. For this thickness, the diffraction efficiency for unpolarized light reaches near unity. This efficiency curve does not include potential reflection and absorption losses. 
              }
         \label{fig:vph_eff_ex}
   \end{figure}

Figure~\ref{fig:vph_eff_ex} shows an example of \textit{s}, and \textit{p}--polarization diffraction efficiency curves as a function of grating thickness $d$.

Tuning a grating to match polarization diffraction efficiency curves thus allows for very high throughput at high dispersion because of the high fringe density (reminding ourselves of Eq.~(\ref{eq:dispersion})). However, it is important to realize that the grating is only fully optimized for one wavelength. Let us call this wavelength $\lambda_{\mathrm{cen}}$. Wavelengths that are different from $\lambda_{\mathrm{cen}}$ will be transmitted in first-order at different diffraction efficiencies. Luckily, Kogelnik also derived an approximation for the full width at half-maximum (FWHM) of the bandwidth ($\Delta \lambda$) due to this effect. This FWHM is given by
\begin{equation}
    \label{eq:FWHM}
    \Delta \lambda \approx \frac{\Lambda}{d} \cot \alpha_2 .
\end{equation}

\begin{figure}
   \centering
   \includegraphics[width=\columnwidth]{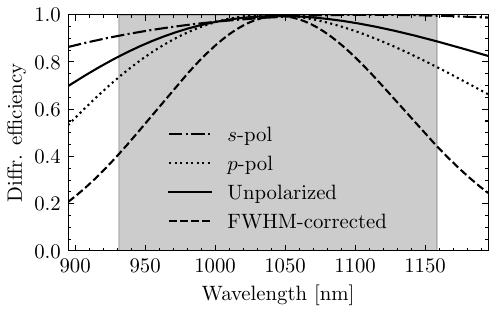}
      \caption{Diffraction efficiency curves versus wavelength. The dotted and dash-dotted lines show the efficiency curves for the two polarization states from Eq.\,(\ref{eq:kogelnik}). The solid line shows the efficiency curve for unpolarized light. The dashed line shows the efficiency curve after having been corrected for the bandwidth using Eq.\,(\ref{eq:FWHM}). The shaded region shows the FWHM.
              }
         \label{fig:new_eff_ex}
   \end{figure}

If we take the optimized grating presented in Fig.~\ref{fig:vph_eff_ex} and take a look at the corrected efficiency curve by multiplying with a Gaussian efficiency function with a FWHM given by Eq.\,(\ref{eq:FWHM}), we find the curve shown in Fig.~\ref{fig:new_eff_ex}. We find that the bandwidth is reduced quite significantly by this effect. Overall, we can conclude that the Dickson grating can be tuned to produce a very high diffraction efficiency in first order but only works over a limited wavelength range.

    \begin{figure}
   \centering
   \includegraphics[width=\columnwidth]{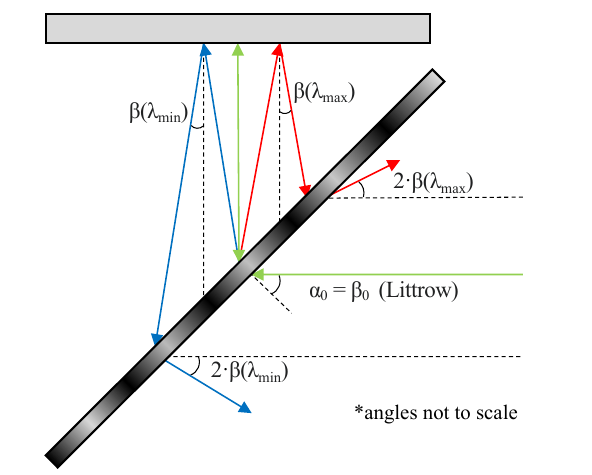}
      \caption{A VPH in double pass. The grating is placed in Littrow, implying that $\alpha_0 = \beta_0$ for $\lambda_{\mathrm{cen}}$. We can see that for wavelengths other than $\lambda_{\mathrm{cen}}$, they will hit the grating at an angle different from $\alpha_0$ at the second pass. How much these rays are off-Littrow is determined by the single-pass dispersion (Eq.~\,\ref{eq:dispersion}).
              }
         \label{fig:dp_diag}
   \end{figure}

       \begin{figure}
   \centering
   \includegraphics[width=\columnwidth]{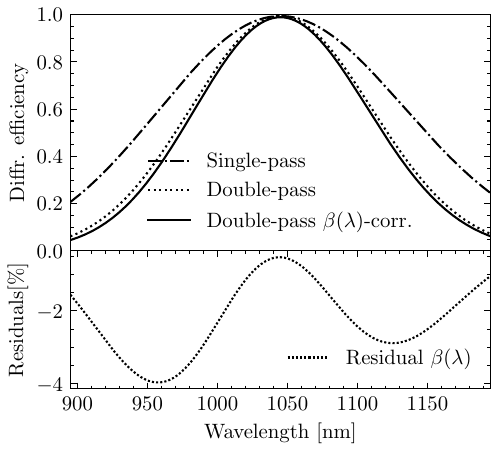}
      \caption{The theoretical diffraction efficiency curves of the grating in single-pass, double-pass (multiplying twice by the Gaussian efficiency function from Fig.~\ref{fig:vph_eff_ex}), and double-pass taking into account the wavelength-dependent incidence angle at the second pass. In the lower frame, we plot the residuals between both double-pass solutions. We find that overall we lose up to $\sim$4 \% in diffraction efficiency when taking into account the wavelength-dependent incidence angle. This fraction is near negligible when comparing it to the efficiency losses caused by the FWHM correction introduced by Eq.~(\ref{eq:FWHM}).
              }
         \label{fig:vph_eff_ex_dp}
   \end{figure}  

\subsection{A VPH in double pass}
In some optical layouts, it can be advantageous to place dispersive (or other) optical elements in a multi-pass configuration. The dispersive power and spectral resolution of that optical element can then be multiplied by a factor equal to the number of passes \citep{wiggins1958, lowenthal1966}. This can reduce the weight, size, complexity, and cost of instruments. For example, putting a VPH grating in double pass will increase the dispersive power and spectral resolution two-fold. However, as the light will pass through the grating twice, this will also reduce the effective bandwidth of the setup. Additionally, as the light is already dispersed at the second pass, this will cause light to hit the grating slightly off-Bragg, reducing the throughput further. However, we will demonstrate that this effect has nearly no impact on our application case.

A diagram of a VPH in double pass can be found in Fig.~\ref{fig:dp_diag}. To determine how the efficiency drops off-Littrow, we can use Eqs~(\ref{eq:dispersion}),~(\ref{eq:kogelnik}). Firstly, we integrate Eq.~(\ref{eq:dispersion}) from $\lambda_{\mathrm{cen}}$ to $\lambda$ to form
\begin{equation}
    \beta(\lambda) = \beta_0 + (\lambda-\lambda_{\mathrm{cen}}) \frac{m \cdot \rho}{\cos\beta_0}.
\end{equation}
Now, at the second pass through the grating, we can define the incident angle as
\begin{equation}
    \alpha_{II,0}(\lambda)=\beta(\lambda).
\end{equation}
Substituting this term into the original Kogelnik efficiency equation (Eq.~\ref{eq:kogelnik}) will give us a wavelength-dependent diffraction efficiency:
\begin{equation}
        \eta(\lambda) = \frac{1}{2}\sin^2 \left( \frac{\pi \Delta n_g d}{\lambda \cos\alpha_{II,0}(\lambda)} \right) +  \frac{1}{2}\sin^2 \left( \frac{\pi \Delta n_g d}{\lambda \cos\alpha_{II,0}(\lambda)} \cos(2\alpha_{II,0}(\lambda)) \right).
\end{equation}

Let us now examine how the new wavelength-dependent diffraction efficiency influences the efficiency curve of our previously simulated grating (Fig.~\ref{fig:vph_eff_ex} and \ref{fig:new_eff_ex}). We find the new theoretical diffraction efficiency curve of the VPH grating in double pass in Fig.~\ref{fig:vph_eff_ex_dp}. Overall we conclude that the effects is minimal and the effective bandwidth with $>50\%$ diffraction efficiency in double pass is $>100$nm.

   \begin{figure}
   \centering
   \includegraphics[width=\columnwidth]{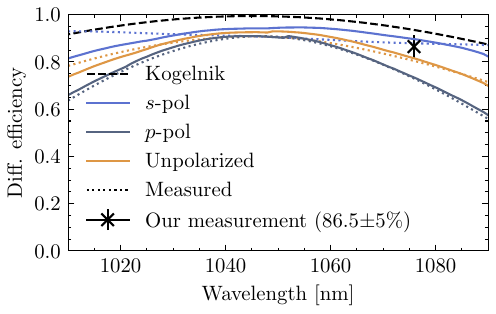}
      \caption{First-order diffraction efficiency measurements and simulations in single-pass configuration of the VPH. The simulated efficiency includes the Kogelnik approximation (black dashed line) and RCWA-derived values. The solid lines (RCWA) and dotted lines (measurement) are supplied by the manufacturer Wasatch Photonics. The measured unpolarized and \textit{p}-polarized efficiency curves demonstrate strong agreement with our RCWA simulated curves. However, the \textit{s}-polarized light measurements show greater deviation. Several factors may contribute to these discrepancies: variations in the bulk refractive index, index modulation and effective thickness of the DCG layer, and to a lesser extent, the thickness and flatness of the AR-coating. These parameters cannot be measured directly, resulting in inevitable small deviations between actual values and those used in our models. These can significantly affect peak efficiency, curve steepness, and centering around the central wavelength, becoming more apparent and having a larger impact towards the bluer end of the spectrum.
              }
         \label{fig:eff_meas}
   \end{figure}

\section{Results}
\subsection{Measurement Setup}

To be able to determine and verify the theoretical performance of the VPH we presented in Sect.~\ref{sect:theory}, we assembled two experimental setups: one to determine diffraction efficiency in single-pass, and one to determine dispersion in double pass. Both measurements relied on many of the same components: a wavelength tunable laser, fiber injection mechanism, spectrum analyzer, collimator, and near-IR detector. Figures~\ref{fig:meas_setup_SP} and \ref{fig:meas_setup_DP} in App.~\ref{app:1} show diagrams of our test setups.

\subsection{Diffraction efficiency}
The wavelength tunable laser used in our measurements is single-mode and tunable from $\sim$1075\,nm to $\sim$1090\,nm. Given the delicate alignment of the rotating setup where we can switch between VPH and fold mirror (see App.~\ref{app:1} Fig.~\ref{fig:meas_setup_SP}), we only performed an efficiency measurement for one wavelength at $1075.9$\,nm. The laser is connected through polarization-maintaining (PM) fibers and attenuators and as such, with how our fiber is injected, we only measured \textit{s}-polarization. Furthermore, depending on the chosen wavelength for our laser, we noticed quick sub-second power oscillations at the photodiode, leading us to believe that the mode-locking of the laser is not very stable. As we will see later in Sect.~\ref{sect:dispersion} this is indeed the case. Likely, the laser mode-switching introduced polarization fluctuations, resulting in power oscillations at the diode. A linear polarizer was added to mitigate any potential polarization crosstalk effects and remove the \textit{p}-polarized component. The measurement wavelength of $1075.9$\,nm was chosen as this was the mode for which we recorded the smallest power fluctuations ($<5\%$). 

In Fig.~\ref{fig:eff_meas} we compare our measured efficiency value to simulated efficiency values, both through the Kogelnik Equation~(Eq.\,(\ref{eq:kogelnik})) as in Fig.~\ref{fig:vph_eff_ex}, rigorous coupled-wave analysis (RCWA) including reflection and absorption losses, and efficiency measurements supplied by the manufacturer. We find that our diffraction efficiency value agrees well with the RCWA simulated values and measurements supplied by the manufacturer. We find that the Kogelnik efficiency value is higher but this is consistent with theory as it does not account for reflection and absorption losses at or in the various layers of the grating. Based on our previously presented simulations in Fig.~\ref{fig:vph_eff_ex_dp}, we expected to be able to reach similar efficiencies at the second pass with a few percent of extra losses at the edges of the band. Overall, looking at the single-pass diffraction efficiency curves in Fig.~\ref{fig:eff_meas}, this grating should allow us to reach $>50\%$ diffraction efficiency over a 100nm bandwidth in double pass, with a peak efficiency around $80\%$. This value will of course depend on the losses at the reflector before the second pass. However, standard silver, aluminum, gold, or dielectric coatings already allow for very high reflection efficiency over a large wavelength regime, so this should not be a major loss-contributor for most wavelengths.  Also, it is important to note that further losses could be induced at the AR coating on second-pass as the incidence angle will not be fully optimal. 

\subsection{Theoretical dispersion and resolution}
\label{sect:dispersiontheory}
As shown by \citet{wiggins1958} \& \citet{lowenthal1966}, for a plane grating in double pass, the dispersion, resolution, and diffraction efficiency should be multiplied by twice that of the single-pass equivalent. They note that this only applies to an ideal case where the grating is in Littrow for both passes and the full beam falls on the grating. Given a grating is typically slightly off-Littrow to offset entrance and exit slits, there are usually some losses. Also, depending on the grating size, it might not allow for the full beam to be captured on the second pass, downgrading efficiency, and possibly resolution. To determine the dispersion and resolution in double pass and verify the hypotheses of \citet{wiggins1958} \& \citet{lowenthal1966} we modified the setup of Fig.~\ref{fig:meas_setup_SP} into the setup shown in Appendix~\ref{app:1} Fig.~\ref{fig:meas_setup_DP}. This setup places the grating in Littrow in single-, and double pass. With the beamsplitter in the second pass, we can isolate the beam that has traveled through the VPH twice. By imaging this spectrum we can verify the dispersive power by changing the laser wavelength and compute angular dispersion from linear dispersion as seen on the detector. The spot size at the focus will tell us the resolving power. 

For our measurements, the wavelength of the laser is tuned to two single modes close to 1077\,nm.
The total beam diameter of the collimator in our setup is 400\,mm. As such, we place an iris in front of the grating to cut the beam to a diameter of 10\,$\pm$\,0.2\,mm. 

We shall express the spectral resolution as the FWHM of a monochromatic line recorded on the detector expressed in linear (as opposite to angular) scale, denoted by parameter $s$. The three components listed below, the entrance slit, iris, and grating, will eventually determine the resolving power, i.e. the wavelength of the line divided by the FWHM of the line expressed in wavelength scale determined by the linear dispersion:
\begin{itemize}
    \item The entrance slit. This is the fiber size (6 $\mu$m monomode fiber in our setup). The spot size on the focal plane is determined by the magnification $M$ (1 in our case), multiplied by the slit size:
    \begin{equation}
        s_{\mathrm{fib}} = d_{\mathrm{fib}}\cdot M = d_{\mathrm{fib}}\cdot\frac{f_{\mathrm{cam}}}{f_{\mathrm{coll}}} = d_{\mathrm{fib}}.
    \end{equation}
    \item The iris. The divergence of the slit is given by $\theta = \frac{\lambda}{D}$. Consequently, the spot size on the focal plane is given by: 
    \begin{equation}
    s_{\mathrm{iris}} = \frac{\lambda}{d_{\mathrm{iris}}}\cdot f_{\mathrm{cam}}.
    \end{equation}
    \item The grating. The reciprocal linear dispersion of the grating is given by
    \begin{equation}
    D_{\mathrm{rec}} = \frac{d_{\mathrm{groove}} \cdot \mathrm{cos}\beta}{2\cdot m \cdot f_{\mathrm{cam}}},
    \end{equation}
    where $d_{\mathrm{groove}}$ is the groove size, $m$ the order of diffraction, and the factor of 2 is introduced by the grating being in double pass, doubling the dispersion and halving the reciprocal dispersion.
    This value has units of nm/mm. As such, it relates a change in wavelength (in nm) across distance on the focal plane (in mm). We can use the equation for the resolving power $R=\lambda / \Delta \lambda$ to derive the spot size: 
    \begin{equation}
    s_{\mathrm{groove}} = \frac{\Delta \lambda}{D_{\mathrm{rec}}} = \frac{2\cdot m \cdot \lambda \cdot f_{\mathrm{cam}}}{R \cdot d_{\mathrm{groove}} \cdot \mathrm{cos}\beta}.
    \end{equation}
    The diffraction-limited resolving power of the double-pass grating is given by $R = 2\cdot N \cdot m$, where $N$ is the number of illuminated lines, given by $d_{\mathrm{iris}}/(\mathrm{cos}\beta\ d_{\mathrm{groove}})$. The multiplication factor of 2 was again added for the double-pass configuration. Consequently, we find that the spot size as a result of divergence from the grating is given by 
    \begin{equation}
    s_{\mathrm{groove}} = \frac{\lambda}{d_{\mathrm{iris}}}\cdot f_{\mathrm{cam}},
    \end{equation}
    equal to the spot size introduced by diffraction from the iris.
\end{itemize}
Trivially, we see that the spot size introduced by divergence from the entrance slit is significantly smaller than the limiting spot size introduced by divergence from our iris and grating. The resolving power of our demonstrator spectrograph should thus be diffraction-limited by the grating/iris, not limited by the size of the entrance slit. Using the equations above we find that the spot size introduced by the iris and grating is $307\pm8$\,$\mu\mathrm{m}$. This means that any spot has a minimum FWHM equal to this value. For a pixel size of 30\,$\mu\mathrm{m}$ this corresponds to roughly 10 pixels. From this FWHM and the reciprocal linear dispersion, as computed before, we can find the maximum resolving power. The smallest resolvable wavelength $\Delta \lambda$ is equal to the FWHM induced by the diffraction limit. In units of wavelength, this is equal to 0.0260\,nm. As such, the maximum achievable resolving power induced by divergence from the iris and grating is:
\begin{equation}
R_\mathrm{max} = \frac{\lambda}{\Delta \lambda} = \frac{1077\ \mathrm{nm}}{0.0260\ \mathrm{nm}} = (4.14\pm0.08)\cdot 10^4.
\end{equation}

In conclusion, for our measurement setup, where the wavelength, camera focal length, and dispersion are fixed, the maximum achievable spectral resolution will always be limited by diffraction from the iris, as long as the grating performs as expected. Given that the maximum achievable spectral resolution induced by the iris aperture and grating are the same, any significant reduction in spectral resolution will be a result from grating imperfections and/or improper alignment of the grating. 

Note that for this calculation and our tests we chose a fairly small iris aperture. Any increase in aperture size would have allowed us to reach higher spectral resolutions. The size of the clear aperture of our grating of $96\times51$\,mm points towards a theoretical upper limit of $R_{\mathrm{max}}=143,000$ for the resolving power. However, at these higher resolutions, the spot size would have become too small for proper sampling with our detector. While changing the camera focal length could have addressed this sampling constraint, it would have prevented us from capturing multiple laser lines within the detector's footprint, which is required to measure the dispersion, as discussed in Sect.~\ref{sect:dispersion}. Therefore a more modest spectral resolution was chosen for this demonstration.

For later comparison, we also determine the (non-reciprocal) linear dispersion $D$. This value solely depends on the camera focal length $f$, the groove density $G$, and the angle of diffraction $\beta$:
\begin{equation}
    \label{eq:lineardisp}
    D=\frac{G\cdot f}{\mathrm{cos \beta}} = \frac{0.0014\ \mathrm{\ell/nm} \cdot 2850\ \mathrm{mm}}{\cos\ 47.5^\circ} = 5.91\ \mathrm{mm/nm}.
\end{equation}
Since our grating is in double pass, the dispersion should be multiplied by a factor of two. The resulting theoretical linear dispersion is thus equal to 11.82\,mm/nm.

\subsection{Measured dispersion and resolution}
\label{sect:dispersion}
\begin{figure}
   \centering
   \includegraphics[width=\columnwidth]{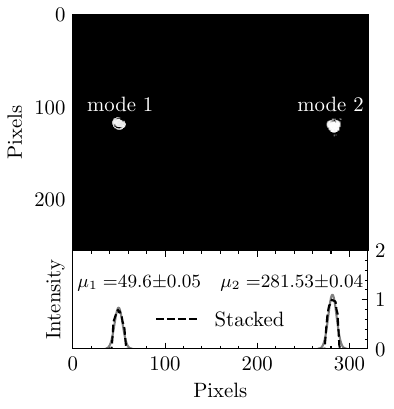}
      \caption{In the upper frame we find two stacked exposures of two different laser modes. The image has been corrected for background flux. We show that the two different modes are spatially separated as a result of dispersion by the VPH. In the lower frame we find the 1D spectrum, resulting from stacking over the vertical axis. The data has been normalized to a peak value of 1 and shown with a dashed line. The solid grey line denotes two Gaussian fits to both peaks. These fits are used to derive the positions on the frame and FWHM of the peaks in order to derive the dispersion and resolving power. $\mu_1$ and $\mu_2$ refer to the mean values of these distributions.
              }
         \label{fig:dp_disp}
   \end{figure}

\begin{figure}
   \centering
   \includegraphics[width=\columnwidth]{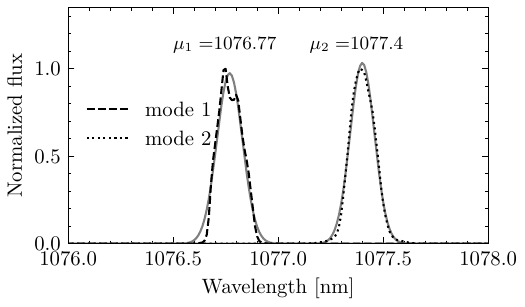}
      \caption{The same two modes as displayed in Fig.~\ref{fig:dp_disp} but now measured with the spectrum analyzer. The dashed and dotted lines show the raw data and the solid grey lines Gaussian fits. The mean values of these distributions are used to derive the wavelength solution of the two modes. 
              }
         \label{fig:spectra_disp}
   \end{figure}

We found two laser modes that were stable in wavelength and within close enough range to both fall on the detector. The resulting spots can be found in Fig.~\ref{fig:dp_disp}. The simultaneous spectrum analyzer measurements can be found in Fig.~\ref{fig:spectra_disp}. Now that we know both the wavelength and spatial separation of the two modes we can compute the dispersion using the collimator focal length and pixel size. The computation is straightforward and can be found in Appendix~\ref{app:2}. We obtain a value of 11.017\,mm/nm, slightly lower than the value derived from theory, which is 11.082\,mm/nm. Very likely this is caused by our grating having been placed slightly off Littrow, changing the dispersion as a consequence. The computation of the real angle of refraction can also be found in App.~\ref{app:2}.

The spectral resolution can be derived from the FWHM of an individual mode as seen on the frame of Fig.~\ref{fig:dp_disp}. This naturally assumes that the intrinsic laser line width is significantly smaller than the FWHM. This is a fair assumption as the laser is a single-mode Fabry Perot laser diode with a cavity length of 3\,mm. Combined with the fact that these are typically gold-coated, resulting in very high cavity finesse, the intrinsic laser line width should be much smaller than the smallest resolvable wavelength (QPhotonics, priv. comm.). We employ the results of the Gaussian fits from above to derive the resolving power. The values are $\sigma_1=4.37\pm0.05\ \mathrm{pix}$ \& $\sigma_2=4.47\pm0.04\ \mathrm{pix}$. Using the pixel size of $30\ \mathrm{\mu m}$ and linear dispersion relation of $11.017\,\mathrm{mm/nm}$ derived in App.~\ref{app:2}, we can convert these values to $\sigma_{\lambda,1}=0.0119\pm0.0002\ \mathrm{nm}$ \& $\sigma_{\lambda,2}=0.0122\pm0.0002\ \mathrm{nm}$. Now to derive the resolving power we employ Eq.~(\ref{eq:Res_sigma}):
\begin{equation}
    \label{eq:Res_sigma}
    \Delta\mathrm{\lambda} = \mathrm{FWHM} = 2 \cdot \sqrt{2\ln(2)}\cdot\sigma_{\lambda} \approx 2.3548 \sigma_{\lambda},
\end{equation}

and Eq.~(\ref{eq:trad_res}):
\begin{equation}
    \label{eq:trad_res}
    R = \frac{\lambda}{\Delta\lambda},
\end{equation}

where $\lambda$ is the wavelength derived from the Gaussian fits in Fig.~\ref{fig:spectra_disp}. This results in resolving powers of $R_1 = (3.84\pm0.05)\cdot 10^4$ \& $R_2 = (3.75\pm0.04)\cdot 10^4$. On average, we obtain a resolving power of $\sim$38k. Adjusting for the lower dispersion than expected from theory, this value corresponds well with our predictions and further supports the claim that our setup is diffraction-limited. Given we do not see a major deviation we can conclude that the grating does not induce any major losses in resolving power. If we had access to more than two single laser lines, we could have improved the precision of our dispersion and resolving power measurements. However, given the very narrow wavelength range studied, we would not expect significant variations within this regime.

\section{Design considerations for a VPH in double pass}

\subsection{The angle-of-incidence and resolving power}
In a non-diffraction-limited case, the maximum resolving power of an unimmersed grating is given by \citep{Baldry2004}:
\begin{equation}
    \label{eq:resolution}
    R = \frac{\lambda}{\Delta\lambda}=\frac{f_{\mathrm{col}}}{\theta_sf_{\mathrm{tel}}} \left(\tan(\alpha_0)+\frac{\sin(\beta_0)}{\cos(\alpha_0)} \right)
\end{equation}

If we take a Littrow case ($\alpha_i=\beta_i$), the equation reduces to the very familiar equation:
\begin{equation}
    \label{eq:resolution2}
    R =\frac{f_{\mathrm{col}}}{\theta_sf_{\mathrm{tel}}} 2\ \tan(\alpha_0)
\end{equation}

\begin{figure}
   \centering
   \includegraphics[width=\columnwidth]{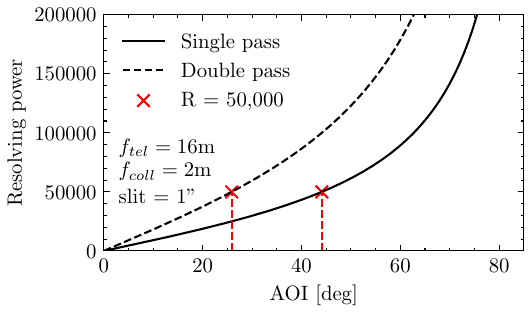}
      \caption{Resolving power versus angle of incidence on an unimmersed grating for a single-pass, and double-pass configuration. The focal lengths and slit size is fixed. We find that at a resolving power of 50,000, the AOI is much less steep for a double-pass configuration than for a single-pass configuration.
              }
         \label{fig:res_aoi}
\end{figure}

For $\alpha_0$ values between $0^\circ$ and $90^\circ$, we find that for a fixed telescope and collimator focal length, and fixed slit size, the resolving power increases for increasing angle following a tangent relation. For a double-pass setup, we already know that we can double the spectral resolving power for the same angle of incidence. In Fig.~\ref{fig:res_aoi} we compare the two trends. Reaching a resolving power of 200,000 would require a very steep AOI of about $75^\circ$ for a single-pass VPH. For a double-pass solution, the AOI is a more modest (but still steep) $63^\circ$. It is important to note here that not every AOI will give high diffraction efficiency. As discussed before, this is only achieved if the Bragg condition is satisfied (Eq.\,(\ref{eq:Bragg})), and this angle is close to the blaze angle of the grating. In other words, a grating that has been designed to work at a specific angle of incidence can typically not be used at a smaller AOI but in double pass to achieve the same resolving power.

Circumventing a very steep AOI would require a smaller slit, typically requiring adaptive optics (AO) to keep high throughput. We can start to see why (for a seeing-limited case) reaching a very high resolving power with a VPH works better for smaller telescopes.

Achieving a steep AOI on the grating is not impossible. For example, one can use slanted fringes and immersion of the grating in prisms (turning it effectively into a GRISM) to achieve a high AOI on the grating but not on the air-to-glass surface. For example \citet{arnsdekker2008} investigated VPH gratings with slanted fringes as a possible alternative for the cross-dispersers in the ESPRESSO instrument \citep{pepe2010} and CODEX \citep{pasquini2010} \citep[which by now goes under the name ANDES,][]{marconi2022}. One could also design the prism after the first pass to use total inner reflection, effectively removing the second air-to-glass interface. However, slanted fringes and GRISMS come with their own caveats, like a changing tilt or curvature of the fringes \citep{rallison1992}, and tight tolerances on optical quality of the prisms \citep{barden2008}. 

\subsection{Tolerances and detector}
\label{sect:tolerances}

For a VPH GRISM, one could utilize either a single block of glass for the prism, which would be extremely heavy given the sizes and angles for high-spectral-resolution gratings, or a set of parallel prisms. The latter option presents challenges in tolerance, as individual prisms would need to be phased in wavelength, introducing manufacturing or assembly tolerances on the order of a fraction of a wavelength. As such, we consider our solution with a plane mirror behind the VPH as the simpler solution. The optical alignment of the mirror is not very challenging and can be compared to the alignment of a blazed grating that is slightly misaligned to offset entrance and exit slits of the spectrograph. Even without a GRISM, operating a VPH in double pass configuration can impose stricter tolerances on the optical quality of its surfaces compared to a single-pass layout ( e.g., on flatness and wavefront-error, WFE ). Especially in a diffraction-limited scenario where the wavefront-error budget is significantly tighter. However, in a diffraction-limited case, gratings can be made smaller to achieve the desired resolution, aiding in reaching tighter tolerances. Consequently, for adaptive-optics-assisted telescopes, where the targets have a smaller footprint in the focal plane, grating footprints can also be smaller. For high-Strehl AO systems, the grating size is independent of the telescope size and solely depends on resolution and wavelength.

\begin{figure}
   \centering
   \includegraphics[width=\columnwidth]{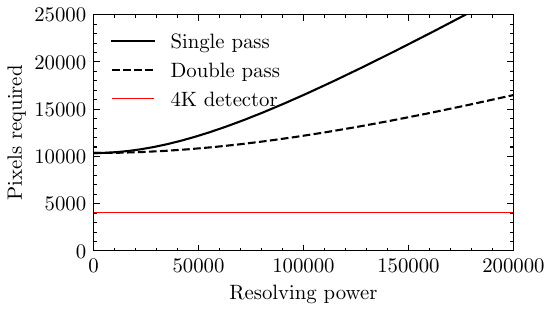}
      \caption{On the horizontal axis we find the resolving power and on the vertical axis the number of pixels required in one line to sample the full effective bandwidth of the grating. Here we used the assumption that every $\Delta\lambda$ element is sampled by two pixels. Two black lines show the curves for a single-, and double pass setup, where the double-pass line is dashed. Note that a single-pass grating requires more pixels because it produces a larger effective bandwidth. A red solid line shows a 4K detector. 
              }
         \label{fig:sampling_resolution}
   \end{figure}

Let us note here that the bandwidth of our VPH spectrograph will be reduced with increasing line density and angle of incidence. If we take a fixed grating period but increasing AOI and thus increasing resolution (see Eq.\,(\ref{eq:resolution})), and compare the number of pixels one would need to sample the full effective bandwidth at 2 pixels per resolution element, we find Fig.~\ref{fig:sampling_resolution}. 

The result is that even 4K detectors are by far not large enough to sample the full efficient bandwidth of these VPHs. This becomes even more significant when one chooses a finer sampling than 2 pixels per resolution element, typically preferred at high spectral resolutions to sample narrow spectral lines. 

Eventually, we point out that this latter discussion is based on a design simplification since a grating cannot be efficient at any AOI (or spectral resolution) for a fixed grating period. Custom optimization will always be necessary to reach maximum efficiency, where one optimizes the grating thickness, refractive index modulation, and groove density to match the Bragg and blaze angle.

When comparing a double-pass VPH with cross-dispersed echelle gratings, we find that the VPH delivers higher efficiencies across the full effective bandwidth measurable by any commercially available detector or even detector mosaics. However, echelle gratings enable coverage of a larger wavelength regime through efficient utilization of 2D detector arrays. Therefore, for applications requiring very broad wavelength coverage, such as high-precision radial velocity measurements where many spectral lines must be resolved, echelle spectroscopy proves more effective. Conversely, for science requiring narrower wavelength bands ($<100\ \mathrm{nm}$), the double-pass VPH offers superior efficiency. Furthermore, in spectrographs with multiple slits or fibers, the VPH presents a practical advantage over echelle gratings by eliminating the risk of overlap between neighboring spectral orders. It should be noted that for VPH spectrographs covering wider wavelength regimes, the camera design necessarily becomes larger, as the extended spectral trace requires larger optics to prevent vignetting.

\section{Conclusions}
We presented a method to achieve (ultra-)high spectral resolution with very high diffraction efficiency by utilizing a VPH grating in a double-pass configuration, as part of the development of the NIGHT instrument. We found that our grating reached a very high double-pass diffraction efficiency of 79\%. Based on manufacturer measurements and RCWA simulations we estimate that it should be possible to reach $>50\%$ double-pass diffraction efficiency over a 100nm bandwidth. Additionally, our findings indicate that the setup is diffraction-limited and that the VPH grating does not compromise spectral resolution when used in a double-pass arrangement. The slight discrepancy between our measured spectral resolution and theoretical expectations can be attributed to a minor misalignment of the grating.

In the field of exoplanet atmospheric characterization, where high spectral resolution over a narrow bandwidth at high signal-to-noise ratio (SNR) is often required, instruments specifically designed around VPH gratings could be extremely beneficial. Operating in the first spectral order results in a large free spectral range, with the primary bandwidth limitation arising from the grating not being fully optimized for a broad range of wavelengths. We show that the bandwidth defined by the grating typically exceeds what can be effectively sampled by standard off-the-shelf detectors. Additionally, we discuss several design considerations that are important for the development of such double-pass VPH-based instruments.

\begin{acknowledgements}
Part of this work has been carried out within the framework of the NCCR PlanetS supported by the Swiss National Science Foundation. We acknowledge support from the Swiss National Science Foundation (SNSF) under grants 184618, 51NF40182901, and 51NF40205606.
We would like to thank the anonymous referee for their constructive feedback that helped improve this manuscript.
\end{acknowledgements}

%
%
\bibliographystyle{aa} 
\bibliography{aa}

\begin{appendix} 
\section{Measurement setups}
\label{app:1}

   \begin{figure*}[h!]
   \centering
   \includegraphics[width=2\columnwidth]{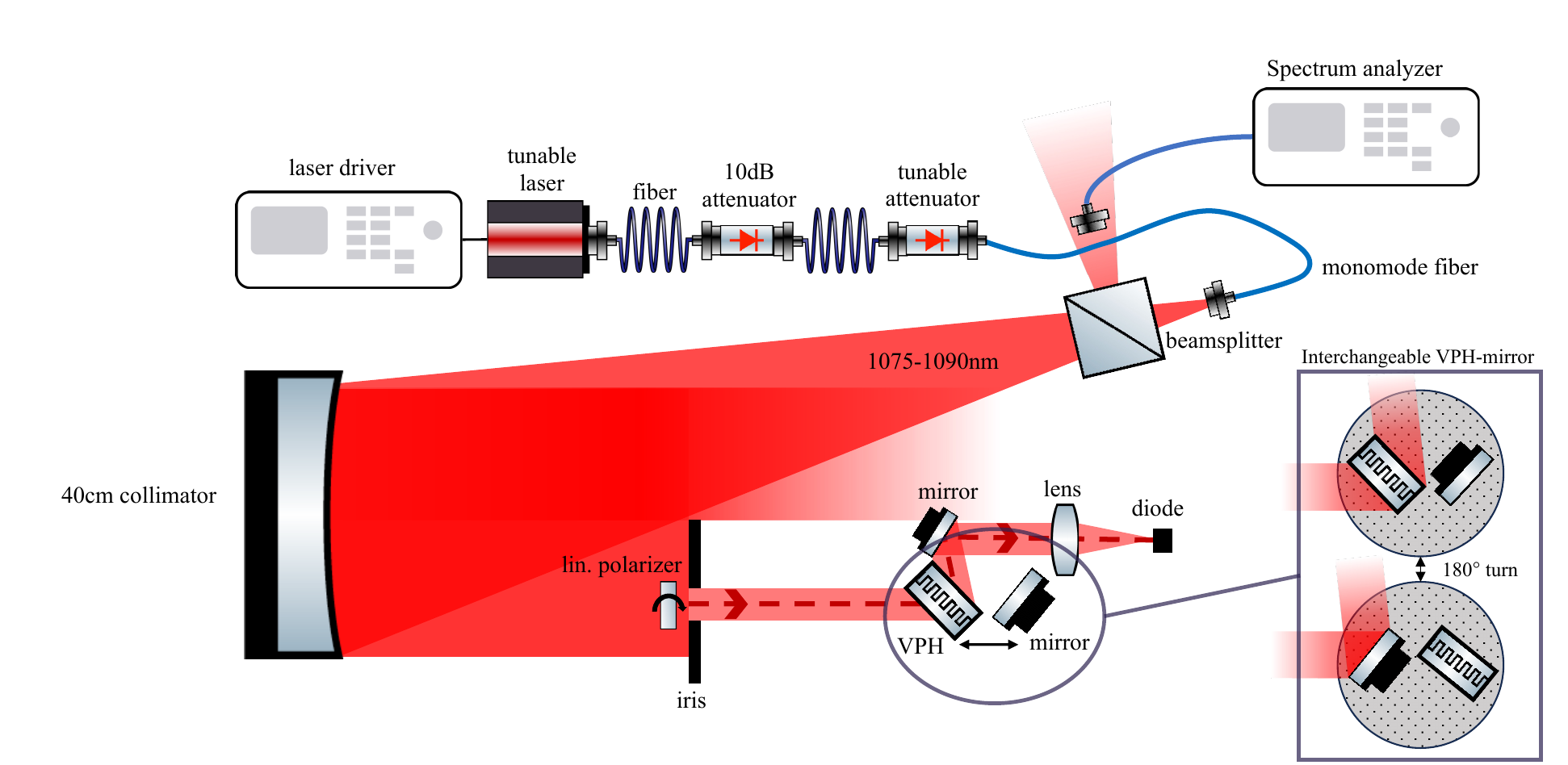}
      \caption{Setup used to determine diffraction efficiency in a single-pass configuration. A laser driver powers a wavelength-tunable laser diode which feeds light into a single-mode (SM) fibre. The fiber is connected to two attenuators, of which one is tunable. These attenuators ensure that the detector is not saturated within reasonable integration times. After the attenuators, light is injected and sent through a 50/50 beamsplitter. The reflected component is sent into a spectrum analyzer which allows for constant monitoring of the wavelength sent out by the tunable laser. The transmitted component is collimated by a 40cm off-axis parabola (focal length: 2580 mm). A linear polarizer and iris form the entrance pupil of the grating measurement setup. The collimated beam can be sent through the VPH or reflected by a plane mirror. A secondary plane fold mirror and a doublet form the camera of the system and focus light on a photodiode which allows for power measurement.
              }
         \label{fig:meas_setup_SP}
   \end{figure*}

   \begin{figure*}[!b]
   \centering
   \includegraphics[width=2\columnwidth]{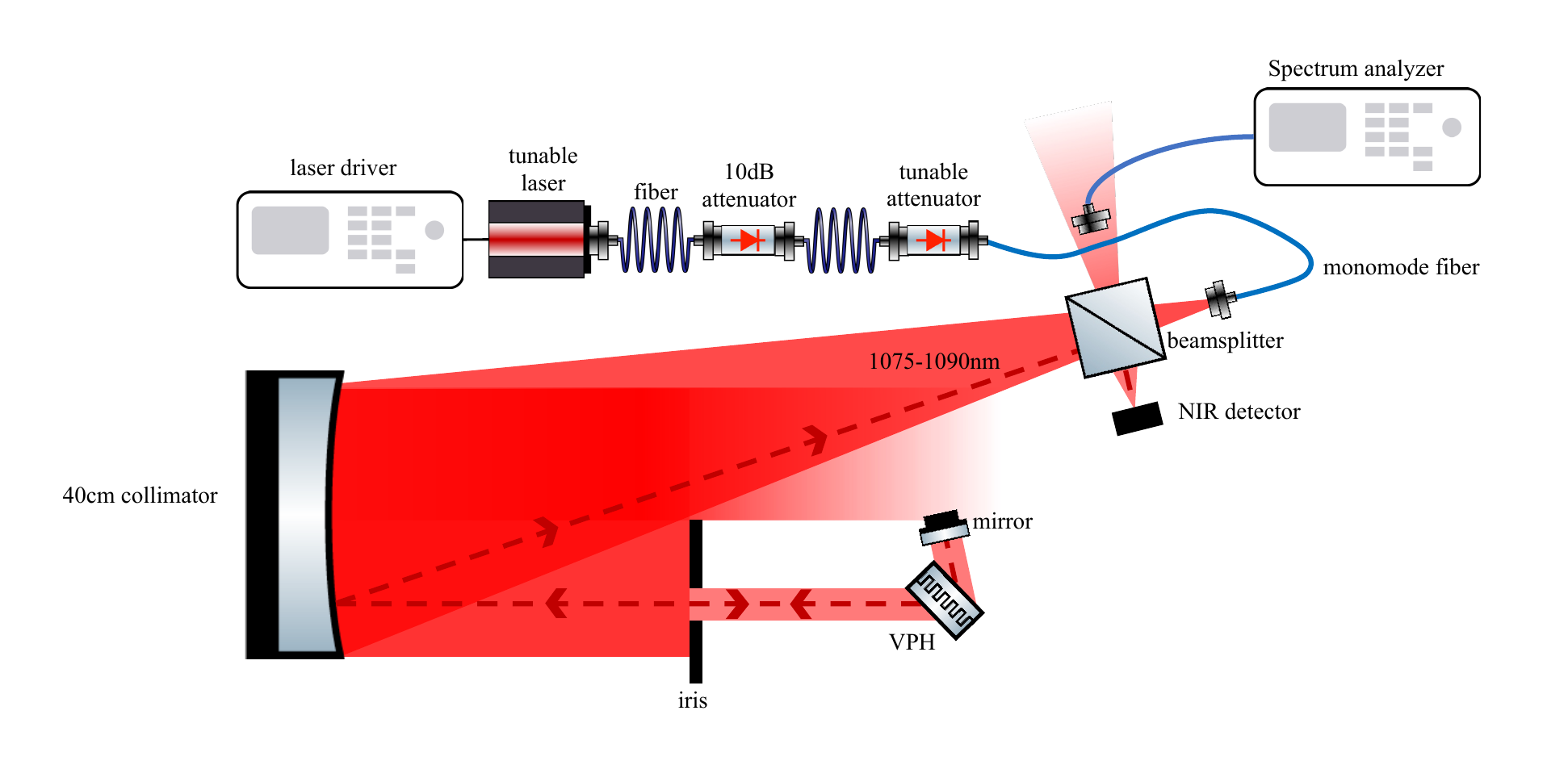}
      \caption{Setup used to determine the dispersion and resolving power of the VPH grating in double pass. Overall, the optical layout is very similar to the setup in Fig.~\ref{fig:meas_setup_SP}. The only difference is that the linear polarizer has been removed and with the help of the fold mirror, the VPH is placed in double pass. The beamsplitter that is used to inject light into the spectrum analyzer is now used in the second pass to re-direct the spectrum to a near-infrared (NIR) detector to image the resulting spectrum.
              }
         \label{fig:meas_setup_DP}
   \end{figure*}

\newpage

\section{Computation of dispersion}
\label{app:2}
The relation between angular deviation $\theta$, collimator length $f$ and spatial separation $H$ follows from basic trigonometry:

\begin{equation*}
    \theta = 2 \cdot \mathrm{tan}^{-1} \left( \frac{H}{2 \cdot f} \right)
\end{equation*}

From the Gaussian fits in Fig.~\ref{fig:dp_disp} we find a pixel separation of 281.53 -- 49.60 = 231.93 pixels. Given a pixel size of 30\,$\mathrm{\mu m}$ this means:
\begin{equation*}
    H/2 = 30\,\mathrm{\mu m} \cdot 231.93 = 6957.9\,\mathrm{\mu m} = 6.9579\,\mathrm{mm}
\end{equation*}

Wavelength separation following from the Gaussian fits in Fig.~\ref{fig:spectra_disp} equals 0.63nm.

This implies that the linear dispersion is equal to:
\begin{equation*}
    \frac{6957.9\,\mathrm{\mu m}}{0.63\,\mathrm{nm}} = 11017\,\mathrm{\mu m}/\mathrm{nm} = 11.017\,\mathrm{mm}/\mathrm{nm}
\end{equation*}

Using the trigonometric relation above we can now compute the angular dispersion:
\begin{equation*}
    \frac{d\beta}{d\lambda} = 2 \cdot \arctan \left( \frac{11.017\,\mathrm{mm/nm}}{2580\,\mathrm{mm}} \right) = 0.00427\,\mathrm{rad/nm}
\end{equation*}

The theoretical linear dispersion computed in Chapter~\ref{sect:dispersiontheory} is 11.82\,mm/nm. Following Eq.\,(\ref{eq:lineardisp}), we can determine the real, measured angle of diffraction $\beta$:

\begin{equation}
    \beta = \arccos \left(\frac{G\cdot f}{D}\right) 
    = \frac{0.0014\,\mathrm{\ell/nm} \cdot 2850\,\mathrm{mm}}{11.017\,\mathrm{mm/nm}} 
    = 47.45^\circ
\end{equation}

\end{appendix}

\end{document}